\documentclass{moriond}

\def\be{\begin{equation}}
\def\ee{\end{equation}}
\def\bea{\begin{eqnarray}}
\def\eea{\end{eqnarray}}

\def\thetal {\ensuremath{\theta_{l}}\xspace}
\def\thetak {\ensuremath{\theta_{K}}\xspace}

\usepackage{ifthen} 
\usepackage{xspace} 
\usepackage{upgreek} 
\newboolean{uprightparticles}
\setboolean{uprightparticles}{false} 
\newboolean{inbibliography}
\setboolean{inbibliography}{false} 

\def\MagUp {\mbox{\em Mag\kern -0.05em Up}\xspace}

\ifthenelse{\boolean{uprightparticles}}%
{

 \def\Pmu         {\ensuremath{\upmu}\xspace}

 \def\Ppi         {\ensuremath{\uppi}\xspace}

 \def\PDelta      {\ensuremath{\Delta}\xspace}                 
 \def\PXi      {\ensuremath{\Xi}\xspace}                 
 \def\PLambda      {\ensuremath{\Lambda}\xspace}                 
 \def\PSigma      {\ensuremath{\Sigma}\xspace}                 
 \def\POmega      {\ensuremath{\Omega}\xspace}                 
 \def\PUpsilon      {\ensuremath{\Upsilon}\xspace}                 
 
 \def\PB      {\ensuremath{\mathrm{B}}\xspace}                 
                  
 \def\PD      {\ensuremath{\mathrm{D}}\xspace}

 \def\PK      {\ensuremath{\mathrm{K}}\xspace}

 \def\Pi      {\ensuremath{\mathrm{i}}\xspace}

 \def\Ps      {\ensuremath{\mathrm{s}}\xspace}

}
{

 \def\Pmu         {\ensuremath{\mu}\xspace}

 \def\Ppi         {\ensuremath{\pi}\xspace}

 \mathchardef\PDelta="7101
 \mathchardef\PXi="7104
 \mathchardef\PLambda="7103
 \mathchardef\PSigma="7106
 \mathchardef\POmega="710A
 \mathchardef\PUpsilon="7107
                  
 \def\PB      {\ensuremath{B}\xspace}                 
                  
 \def\PD      {\ensuremath{D}\xspace}

 \def\PK      {\ensuremath{K}\xspace}

 \def\Pi      {\ensuremath{i}\xspace}

 \def\Ps      {\ensuremath{s}\xspace}

}

\makeatletter
\ifcase \@ptsize \relax
  \newcommand{\miniscule}{\@setfontsize\miniscule{4}{5}}
\or
  \newcommand{\miniscule}{\@setfontsize\miniscule{5}{6}}
\or
  \newcommand{\miniscule}{\@setfontsize\miniscule{5}{6}}
\fi
\makeatother

\DeclareRobustCommand{\optbar}[1]{\shortstack{{\miniscule (\rule[.5ex]{1.25em}{.18mm})}
  \\ [-.7ex] $#1$}}

\def\mup        {{\ensuremath{\Pmu^+}}\xspace}
\def\mun        {{\ensuremath{\Pmu^-}}\xspace} 
\def\mumu       {{\ensuremath{\Pmu^+\Pmu^-}}\xspace}

\def\squark    {{\ensuremath{\Ps}}\xspace}

\def\pion   {{\ensuremath{\Ppi}}\xspace}

\def\pip    {{\ensuremath{\pion^+}}\xspace}

\def\kaon    {{\ensuremath{\PK}}\xspace}
  \def\Kbar    {{\kern 0.2em\overline{\kern -0.2em \PK}{}}\xspace}

\def\KorKbar    {\kern 0.18em\optbar{\kern -0.18em K}{}\xspace}

\def\Kp      {{\ensuremath{\kaon^+}}\xspace}
\def\Km      {{\ensuremath{\kaon^-}}\xspace}

\def\Kstarz  {{\ensuremath{\kaon^{*0}}}\xspace}

  \def\Dbar    {{\kern 0.2em\overline{\kern -0.2em \PD}{}}\xspace}

\def\DorDbar    {\kern 0.18em\optbar{\kern -0.18em D}{}\xspace}

\def\B       {{\ensuremath{\PB}}\xspace}
\def\Bbar    {{\ensuremath{\kern 0.18em\overline{\kern -0.18em \PB}{}}}\xspace}

\def\BorBbar    {\kern 0.18em\optbar{\kern -0.18em B}{}\xspace}

\def\Bu      {{\ensuremath{\B^+}}\xspace}

\def\Bd      {{\ensuremath{\B^0}}\xspace}
\def\Bs      {{\ensuremath{\B^0_\squark}}\xspace}

  \def\Y#1S{\ensuremath{\PUpsilon{(#1S)}}\xspace}

\def\Lbar        {{\ensuremath{\kern 0.1em\overline{\kern -0.1em\PLambda}}}\xspace}
\def\LorLbar    {\kern 0.18em\optbar{\kern -0.18em \PLambda}{}\xspace}

\newcommand{\decay}[2]{\ensuremath{#1\!\to #2}\xspace}         

\def\to                 {\ensuremath{\rightarrow}\xspace}

\def\CP                {{\ensuremath{C\!P}}\xspace}

\def\AT#1     {\ensuremath{A_{\mathrm{T}}^{#1}}\xspace}           

\def\C#1      {\ensuremath{\mathcal{C}_{#1}}\xspace}                       
\def\Cp#1     {\ensuremath{\mathcal{C}_{#1}^{'}}\xspace}                    
\def\Ceff#1   {\ensuremath{\mathcal{C}_{#1}^{\mathrm{(eff)}}}\xspace}        
\def\Cpeff#1  {\ensuremath{\mathcal{C}_{#1}^{'\mathrm{(eff)}}}\xspace}       
\def\Ope#1    {\ensuremath{\mathcal{O}_{#1}}\xspace}                       
\def\Opep#1   {\ensuremath{\mathcal{O}_{#1}^{'}}\xspace}                    

\newcommand{\tev}{\ifthenelse{\boolean{inbibliography}}{\ensuremath{~T\kern -0.05em eV}\xspace}{\ensuremath{\mathrm{\,Te\kern -0.1em V}}}\xspace}
\newcommand{\gev}{\ensuremath{\mathrm{\,Ge\kern -0.1em V}}\xspace}
\newcommand{\mev}{\ensuremath{\mathrm{\,Me\kern -0.1em V}}\xspace}
\newcommand{\kev}{\ensuremath{\mathrm{\,ke\kern -0.1em V}}\xspace}
\newcommand{\ev}{\ensuremath{\mathrm{\,e\kern -0.1em V}}\xspace}
\newcommand{\gevc}{\ensuremath{{\mathrm{\,Ge\kern -0.1em V\!/}c}}\xspace}
\newcommand{\mevc}{\ensuremath{{\mathrm{\,Me\kern -0.1em V\!/}c}}\xspace}
\newcommand{\gevcc}{\ensuremath{{\mathrm{\,Ge\kern -0.1em V\!/}c^2}}\xspace}
\newcommand{\gevgevcccc}{\ensuremath{{\mathrm{\,Ge\kern -0.1em V^2\!/}c^4}}\xspace}
\newcommand{\mevcc}{\ensuremath{{\mathrm{\,Me\kern -0.1em V\!/}c^2}}\xspace}

\def\invfb   {\ensuremath{\mbox{\,fb}^{-1}}\xspace}

\def\deriv {\ensuremath{\mathrm{d}}}

\def\gsim{{~\raise.15em\hbox{$>$}\kern-.85em
          \lower.35em\hbox{$\sim$}~}\xspace}
\def\lsim{{~\raise.15em\hbox{$<$}\kern-.85em
          \lower.35em\hbox{$\sim$}~}\xspace}

\def\tell1  {TELL1\xspace}
\def\ukl1   {UKL1\xspace}

\newcommand{\Photo}{}

\begin{document}
\vspace*{4cm}
\title{Rare decays at LHCb}

\author{C.\ Langenbruch$^{1,2}$ on behalf of the LHCb collaboration}

\address{$^1$University of Warwick, Department of Physics, Gibbet Hill Road, Coventry CV4\,7AL, UK\\
$^2$RWTH Aachen, I.\ Physikalisches Institut B, Sommerfeldstr.\ 14, 52074 Aachen, Germany  
}

\maketitle\abstracts{
  Rare decays are flavour changing neutral current processes that are 
  loop-suppressed in the Standard Model (SM). 
  New particles in SM extensions can therefore give significant contributions,
  modifying branching fractions and angular distributions. 
  Consequently, rare decays are particularly sensitive probes for New Physics (NP).
  These proceedings summarize the latest results from the LHCb experiment on rare decays. 
  While most results are in good agreement with SM predictions, some tensions that recently appeared 
  in rare semileptonic $\decay{b}{s\ell^+\ell^-}$ decays are also discussed. 
}

\section{Introduction}
Rare decays that proceed via $b\to s(d)$ quark level transitions constitute flavour changing neutral currents and, in the SM, are forbidden at tree-level and can only occur at loop-level.
Contributions from NP
can therefore be comparably large 
and significantly affect branching fractions and angular distributions. 
Precision measurements of rare decays therefore constitute sensitive searches for NP.
Furthermore, they allow to determine the underlying operator structure of potential new contributions in global fits~\cite{Altmannshofer:2014rta,Descotes-Genon:2015uva,Hurth:2016fbr,Beaujean:2013soa}.
These proceedings present recent results on rare decays from LHCb, determined using $3\invfb$ of data taken during the LHC Run~1. 

\section{Observation of the very rare decay $\decay{\Bs}{\mumu}$}
The very rare decay $\decay{\Bs}{\mumu}$ is not only loop but also helicity suppressed. 
Due to the fully leptonic final state, the decay is experimentally well accessible and theory predictions are particularly precise. 
The SM prediction for the branching fraction of the decay is given in Ref.~\cite{Bobeth:2013uxa} as ${\cal B}(\decay{\Bs}{\mumu})_{\rm SM}=(3.66\pm 0.23)\times 10^{-9}$. 
The related decay $\decay{\Bd}{\mumu}$ is further suppressed by the ratio of CKM matrix elements $\left|V_{\rm td}/V_{\rm ts}\right|^2$, resulting in a SM prediction of ${\cal B}(\decay{\Bd}{\mumu})_{\rm SM}=(1.06\pm 0.09)\times 10^{-10}$. 
The decays are particularly sensitive to contributions from NP in the (pseudo)scalar sector, since these possible new contributions are 
enhanced compared to the SM contribution.
The ratio of the branching fractions of the two decays is ${\cal R}={\cal B}(\decay{\Bd}{\mumu})_{\rm SM}/{\cal B}(\decay{\Bs}{\mumu})_{\rm SM}=0.0295^{+0.0028}_{-0.0025}$ for the SM, as well as in NP models with the property of minimal flavour violation. 

\begin{figure}
  \centering
  \includegraphics[height=4.3cm]{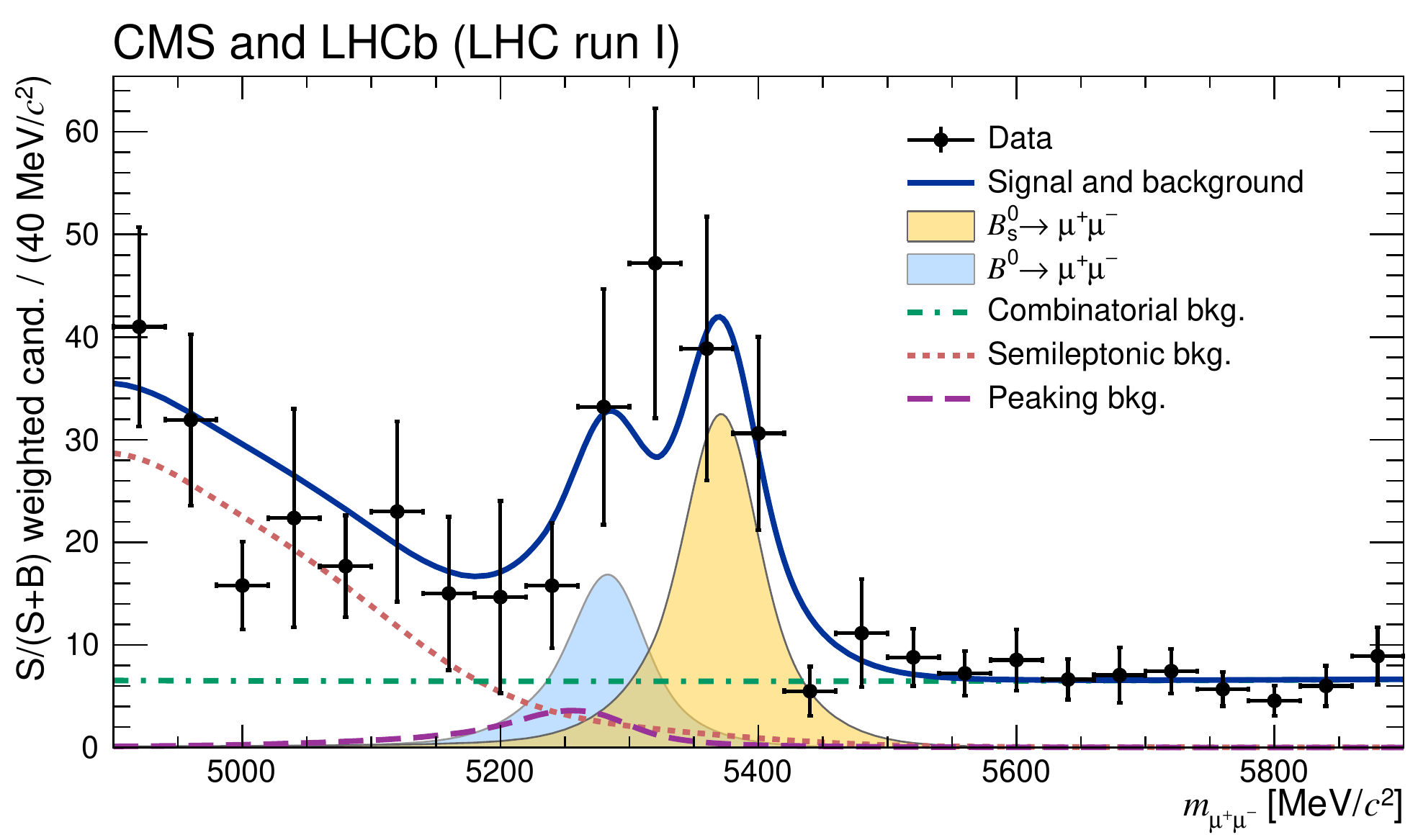}
  \includegraphics[height=4.2cm]{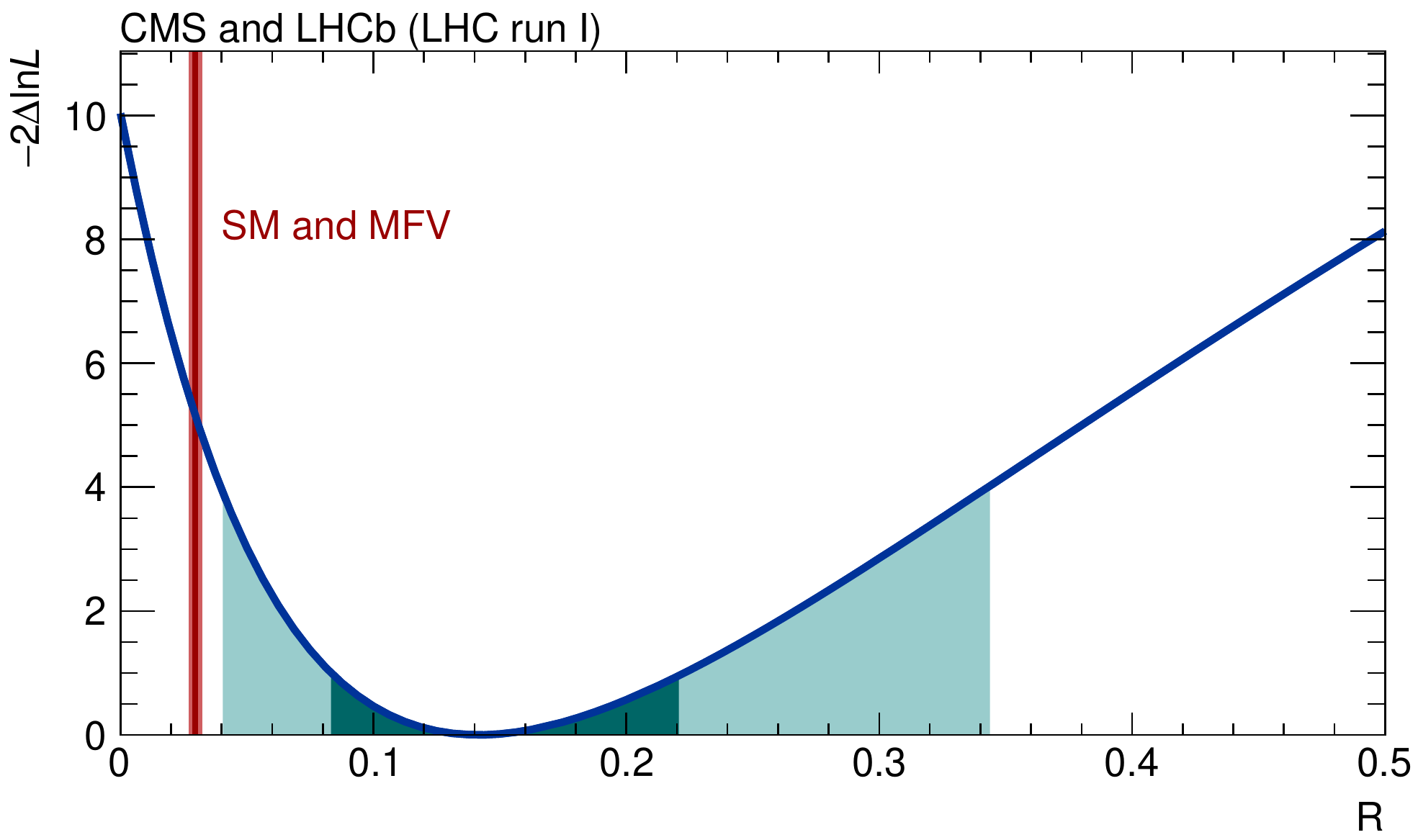}  
  \caption{(Left) $\decay{B^0_{(s)}}{\mumu}$ candidates in the combined CMS and LHCb dataset.
    (Right) branching fraction ratio ${\cal R}={\cal B}(\decay{\Bd}{\mumu})/{\cal B}(\decay{\Bs}{\mumu})$.
    The dark (light) green regions define the $1\,\sigma$ ($2\,\sigma$) confidence intervals.\label{fig:bsmumu}}
\end{figure}

The CMS and LHCb collaborations have performed a combined analysis~\cite{CMS:2014xfa}. 
Figure~\ref{fig:bsmumu} (left) shows the signal candidates overlaid with a combined fit sharing signal and nuisance parameters. 
The measured branching fractions are
${\cal B}(\decay{\Bs}{\mumu}) = (2.8^{+0.7}_{-0.6})\times 10^{-9}$ and ${\cal B}(\decay{\Bd}{\mumu}) = (3.9^{+1.6}_{-1.4})\times 10^{-10}$,
in agreement with the SM prediction at $1.2\,\sigma$ and $2.2\,\sigma$, respectively. 
The decay $\decay{\Bs}{\mumu}$ is observed with a significance of $6.2\,\sigma$. 
First evidence is found for the decay $\decay{\Bd}{\mumu}$ with a significance of $3.0\,\sigma$. 
The branching fraction ratio is ${\cal R}=0.14^{+0.08}_{-0.06}$ and is shown in Fig.~\ref{fig:bsmumu} (right).
It is compatible with the SM and minimal flavour violation at $2.3\,\sigma$. 

\section{Rare electroweak penguin decays}
\subsection{The decay $\decay{\Bd}{\Kstarz\mumu}$}
The rare decay $\decay{\Bd}{\Kstarz\mumu}$ gives access to many angular observables that are sensitive to NP contributions.
The decay is fully defined by three decay angles $\vec{\Omega}=(\thetal, \thetak, \phi)$, and $q^2$, the invariant mass of the dimuon system squared. 
The \CP-averaged angular distribution in a bin of $q^2$ is given by~\cite{Altmannshofer:2008dz}
\begin{eqnarray*}
\frac{1}{\deriv(\Gamma+\bar{\Gamma})/\deriv q^2}\frac{\deriv^3(\Gamma+\bar{\Gamma})}{\deriv\vec{\Omega}} =
\frac{9}{32\pi} &\Big[
 & \frac{3}{4} (1-{F_{\rm L}})\sin^2\thetak + {F_{\rm L}}\cos^2\thetak 
+ \frac{1}{4}(1-{F_{\rm L}})\sin^2\thetak\cos 2\thetal\nonumber\\
&-& {F_{\rm L}} \cos^2\thetak\cos 2\thetal + {S_3}\sin^2\thetak \sin^2\thetal \cos 2\phi\nonumber\\
&+& {S_4} \sin 2\thetak \sin 2\thetal \cos\phi + {S_5}\sin 2\thetak \sin \thetal \cos \phi\nonumber\\
&+& \frac{4}{3} {A_{\rm FB}} \sin^2\thetak \cos\thetal + {S_7} \sin 2\thetak \sin\thetal \sin\phi\nonumber\\
&+& {S_8} \sin 2\thetak \sin 2\thetal \sin\phi + {S_9}\sin^2\thetak \sin^2\thetal \sin 2\phi  \Big],
\end{eqnarray*}
where $F_{\rm L}$ denotes the longitudinal polarisation fraction of the $\Kstarz$ and $A_{\rm FB}$ the forward-backward asymmetry of the dimuon system. 
In Ref.~\cite{Descotes-Genon:2013vna} an alternative parametrisation using the $P_i^{(\prime)}$ observables was proposed, that are designed such that hadronic form-factor uncertainties cancel at leading order. 
An example is the observable $P_5^\prime$, defined as $P_5^\prime=S_5/\sqrt{F_{\rm L}(1-F_{\rm L})}$. 

\begin{figure}
  \centering
  \includegraphics[height=4.3cm]{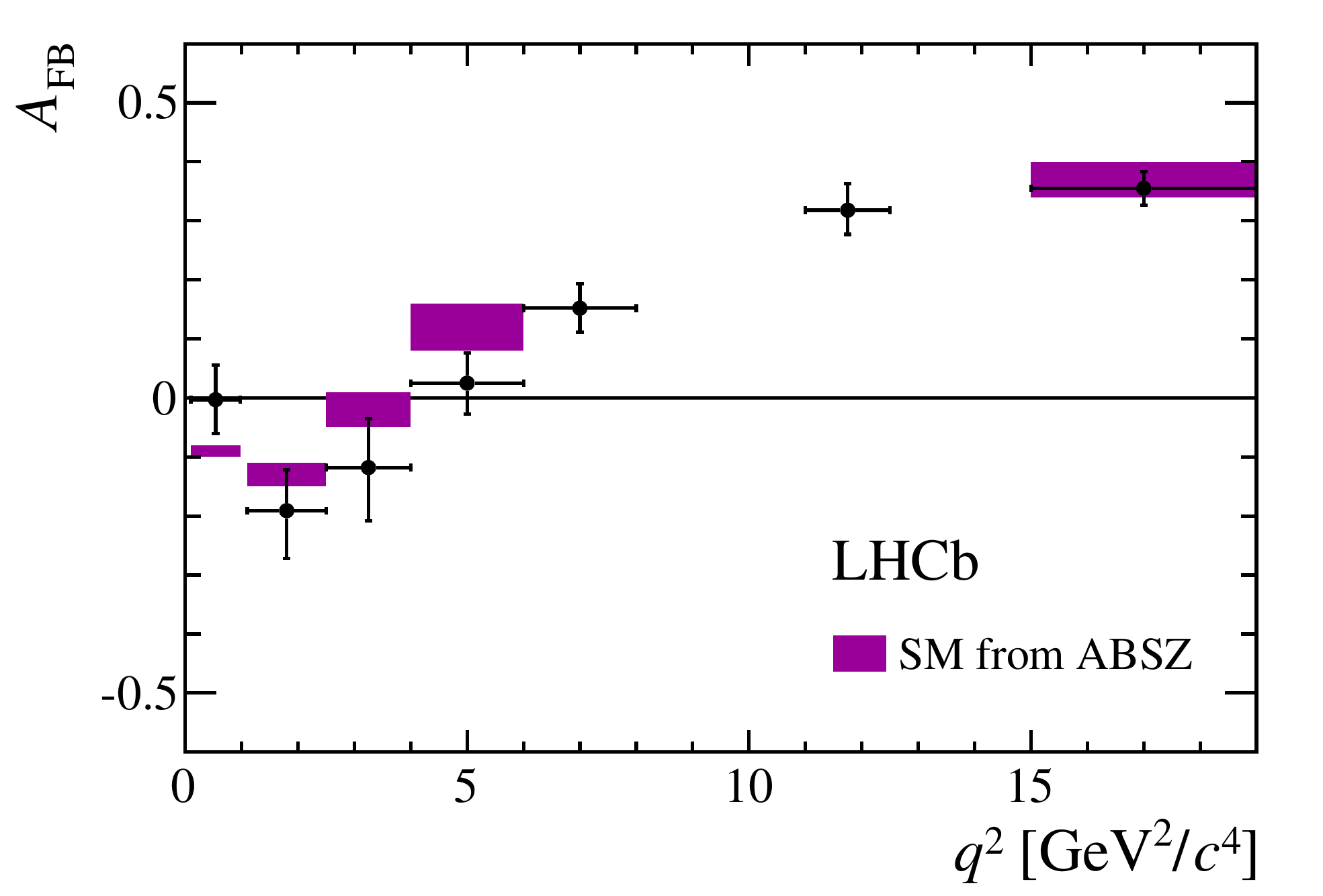}
  \includegraphics[height=4.3cm]{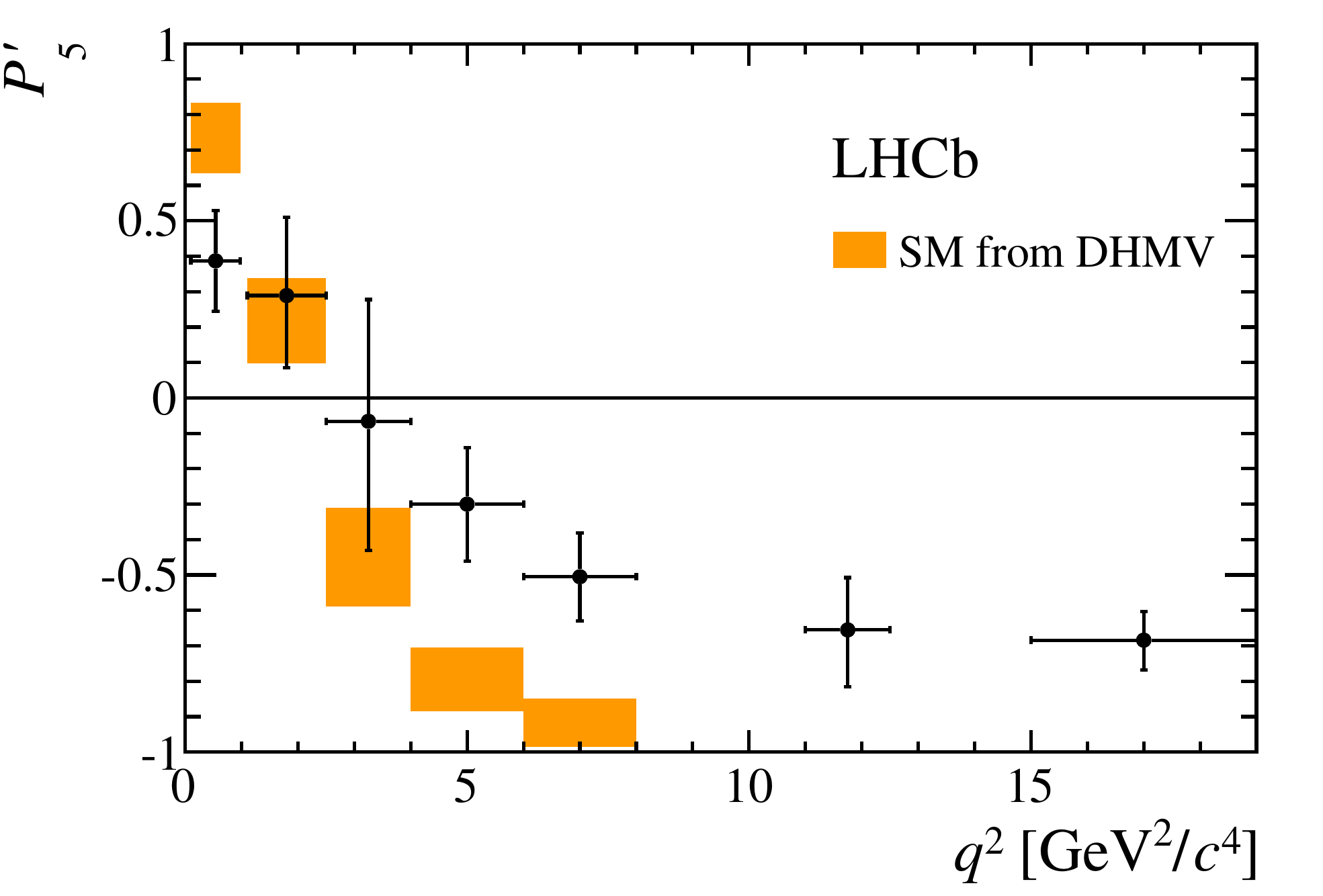}
  \caption{
    (Left) The forward-backward asymmetry $A_{\rm FB}$,
    overlaid with SM predictions from Refs.~\protect\cite{Altmannshofer:2014rta,Straub:2015ica}. 
    (Right) The angular observable $P_5^\prime$, 
    overlaid with SM predictions from Ref.~\protect\cite{Descotes-Genon:2014uoa}. 
    \label{fig:kstarmumu}}
\end{figure}

The LHCb collaboration performed the first full angular analysis of the decay $\decay{\Bd}{\Kstarz\mumu}$~\cite{Aaij:2015oid}. 
Figure~\ref{fig:kstarmumu} (left) shows $A_{\rm FB}$ overlaid with 
SM predictions from Refs.~\cite{Altmannshofer:2014rta,Straub:2015ica}. 
In the $q^2$ region $1.1<q^2<6.0\gevgevcccc$, the data lies below the prediction, but overall good agreement is observed. 
The observable $P_5^\prime$ is given in Fig.~\ref{fig:kstarmumu} (right), together with the SM prediction from Ref.~\cite{Descotes-Genon:2014uoa}. 
In the two $q^2$ bins $[4,6]\gevgevcccc$ and $[6,8]\gevgevcccc$ local deviations from the SM prediction are observed that correspond to $2.8\,\sigma$ and $3.0\,\sigma$. 
A global analysis of all \CP-averaged $\decay{\Bd}{\Kstarz\mumu}$ observables finds a global deviation of $3.4\,\sigma$ from the SM prediction. 
The \CP\ asymmetries $A_{3,\ldots,9}$ are also measured and show good agreement with the SM expectation. 

\subsection{The decay $\decay{\Bs}{\phi\mumu}$}
\begin{figure}
  \centering
  \includegraphics[height=4.3cm]{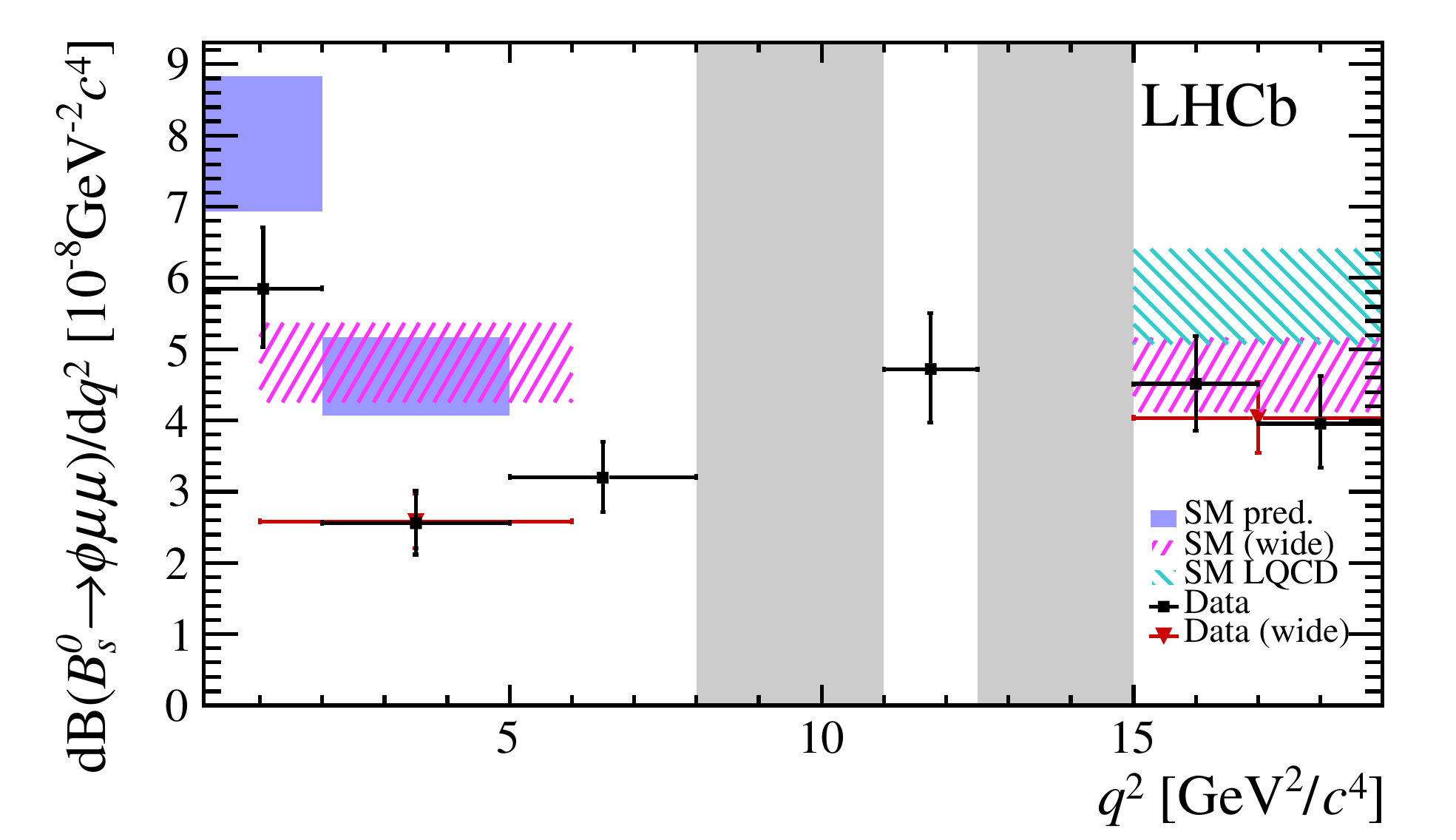}  
  \includegraphics[height=4.5cm]{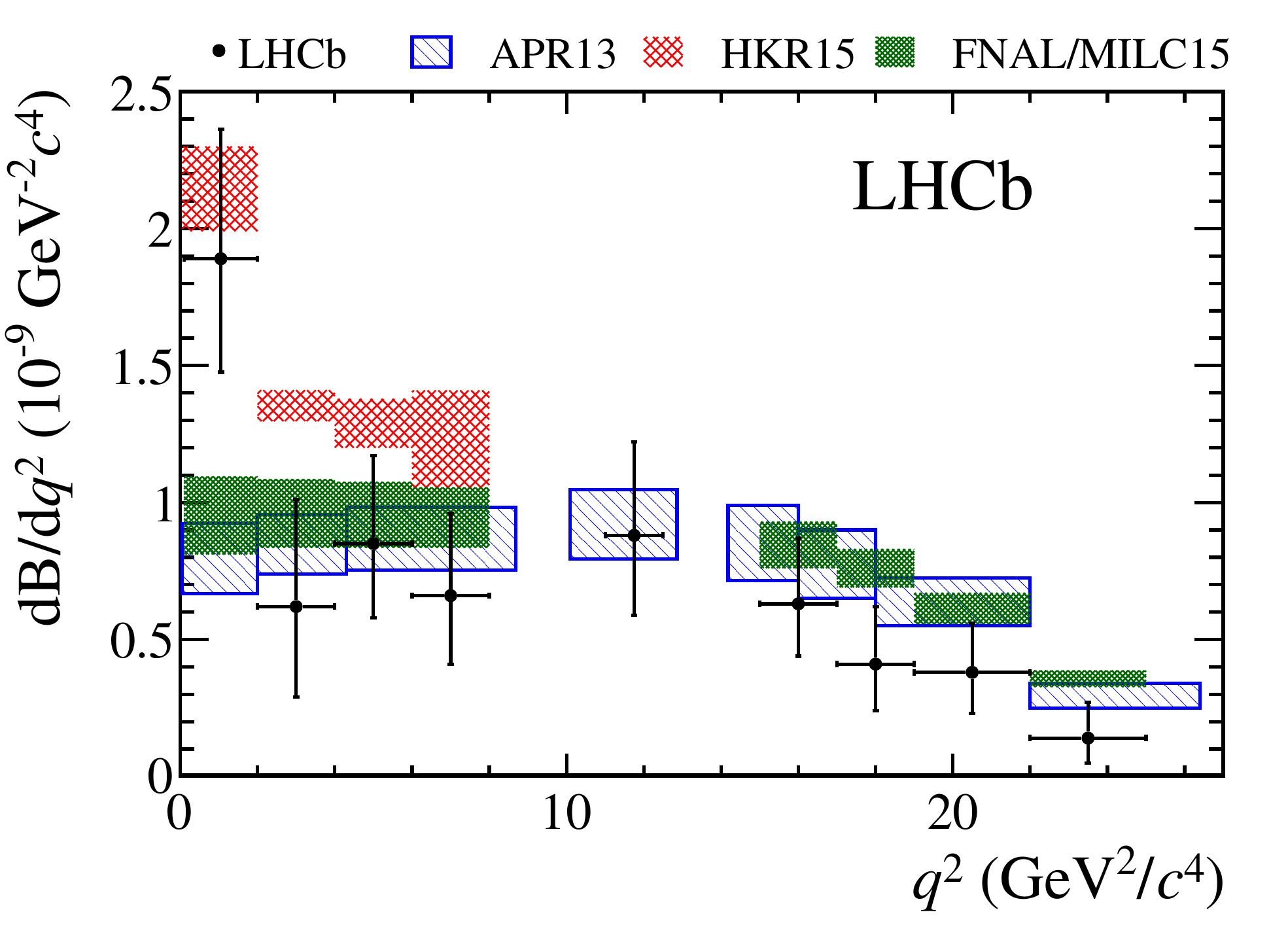}
  \caption{The differential branching fraction of (left) the decay $\decay{\Bs}{\phi\mumu}$,
    overlaid with SM predictions from Refs.~\protect\cite{Altmannshofer:2014rta,Straub:2015ica,Horgan:2013pva} and 
    (right) the decay $\decay{\Bu}{\pip\mumu}$,
    overlaid with SM predictions from Refs.~\protect\cite{Ali:2013zfa,Hambrock:2015wka,Bailey:2015nbd}.
    \label{fig:diffbfs}}
\end{figure}

The decay $\decay{\Bs}{\phi\mumu}$ is the dominant $\decay{b}{s\mumu}$ decay in the $\Bs$ system. 
The fact that the final state $\phi(\to\Kp\Km)\mup\mun$ is not flavour specific reduces the number of angular observables accessible in this decay compared to 
the decay $\decay{\Bd}{\Kstarz\mumu}$. 
LHCb performed a full angular analysis and measurement of the differential branching faction~\cite{Aaij:2015esa}. 
The angular observables are found to be in good agreement with SM predictions. 
Figure~\ref{fig:diffbfs} (left) shows the differential branching fraction, overlaid with the SM predictions from Refs.~\cite{Altmannshofer:2014rta,Straub:2015ica,Horgan:2013pva}. In the low $q^2$ region $1<q^2<6\gevgevcccc$ the differential branching fraction is found to be $3.3\,\sigma$ below the SM prediction. 

\subsection{The decay $\decay{\Bu}{\pip\mumu}$}
The decay $\decay{\Bu}{\pip\mumu}$ is a $b\to d\mumu$ transition and therefore in the SM suppressed by $\left|V_{\rm td}/V_{\rm ts}\right|^2$ compared to the corresponding $\decay{b}{s\mumu}$ decay $\decay{\Bu}{\Kp\mumu}$.
LHCb has performed a measurement of the differential branching fraction and the \CP\ asymmetry of this decay~\cite{Aaij:2015nea}. 
Figure~\ref{fig:diffbfs} (right) shows the differential branching fraction, overlaid with SM predictions from Refs.~\cite{Ali:2013zfa,Hambrock:2015wka,Bailey:2015nbd}. 
Good agreement of the data with the SM predictions is observed, which is further improved in Ref.~\cite{Hambrock:2015wka} where contributions from light resonances are included. 
The \CP\ asymmetry of the decay is measured to be ${\cal A}_{\CP}=-0.11\pm 0.12\pm 0.01$, in good agreement with the SM expectation. 

In addition, the size of the CKM matrix elements $|V_{\rm td}|$ and $|V_{\rm ts}|$ as well as their ratio are determined as 
$|V_{\rm td}|=7.2^{+0.9}_{-0.8}\times 10^{-3}$, $|V_{\rm ts}|=3.2^{+0.4}_{-0.4}\times 10^{-2}$ and $|V_{\rm td}/V_{\rm ts}|=0.24^{+0.05}_{-0.04}$. 
Recent form-factors from lattice calculations allow for a determination of the CKM matrix elements from 
rare decays~\cite{Aaij:2015nea,Aaij:2014pli} 
with a precision similar to the determination from mixing measurements~\cite{Du:2015tda,Bazavov:2016nty}. 

\section{Lepton universality tests using rare decays}
The ratio 
${\cal R}_K(q^2_{\rm min}, q^2_{\rm max}) = \bigl[\int_{q^2_{\rm min}}^{q^2_{\rm max}} {\rm d}{\cal B}(\decay{\Bu}{\Kp\mumu})/{\rm d}q^2\bigr]/\bigl[\int_{q^2_{\rm min}}^{q^2_{\rm max}} {\rm d}{\cal B}(\decay{\Bu}{\Kp e^+e^-})/{\rm d}q^2\bigr]$ 
is a sensitive test of lepton universality.
In the SM, the value of ${\cal R}_K$ in the range $1<q^2<6\gevgevcccc$ is precisely predicted to be $1\pm {\cal O}(10^{-3})$~\cite{Bobeth:2007dw}. Hadronic uncertainties largely cancel in the ratio. 
The LHCb collaboration finds ${\cal R}_K=0.745^{+0.090}_{-0.074}\pm 0.036$, 
in tension with the SM prediction at $2.6\,\sigma$~\cite{Aaij:2014ora}. 
Further lepton universality tests at LHCb are in preparation, including measurements of ${\cal R}_{K^*}$ and ${\cal R}_\phi$.
Furthermore, the ${\cal R}_K$ measurement motivates searches for lepton flavour violating decays~\cite{Hiller:2014yaa,Glashow:2014iga}. 

\section{Conclusions}
Recent results on rare decays from LHCb have been presented. 
While most are in good agreement with SM predictions, some tensions exist, most notably in the angular observables of the decay $\decay{\Bd}{\Kstarz\mumu}$,
the branching fraction of $\decay{\Bs}{\phi\mumu}$ and ${\cal R}_K$. 
Global fits of the data on rare $b\to s$ transitions indicate 
a significance of this tension of around $3$--$4\,\sigma$~\cite{Altmannshofer:2014rta,Descotes-Genon:2015uva}. 
While consistent NP explanations for the deviations in the form of new heavy gauge bosons~\cite{Glashow:2014iga,Gauld:2013qba,Buras:2013qja,Crivellin:2015mga} and leptoquarks~\cite{Hiller:2014yaa,Biswas:2014gga,Buras:2014fpa,Gripaios:2014tna} have been discussed, 
underestimated hadronic uncertainties cannot be excluded~\cite{Jager:2014rwa,Lyon:2014hpa,Ciuchini:2015qxb}. 

The presented measurements motivate further work both in theory as well as experiment. 
With the upcoming Run~2 data, LHCb will perform further analyses of rare $b\to s$ decays,
including additional tests of lepton universality and searches for lepton flavour violating decays.

\end{document}